\documentclass[10pt,twocolumn,english,aps,pra,superscriptaddress,address,showpacs,showacknowledgments]{revtex4-1}
\usepackage[T1]{fontenc}
\usepackage[latin9]{inputenc}
\usepackage[a4paper]{geometry}
\usepackage[active]{srcltx}
\usepackage{xcolor}
\usepackage{amsmath}
\usepackage{amssymb}
\usepackage{stackrel}
\usepackage{graphicx}
\PassOptionsToPackage{normalem}{ulem}
\usepackage{ulem}

\makeatletter

\providecolor{lyxadded}{rgb}{0,0,1}
\providecolor{lyxdeleted}{rgb}{1,0,0}

\DeclareRobustCommand{\lyxdeleted}[3]{{\color{lyxdeleted}\sout{#3}}}

\usepackage{babel}
\begin{document}

\title{Two-fold mechanical squeezing in a cavity optomechanical system}

\author{Chang-Sheng Hu}

\affiliation{Fujian Key Laboratory of Quantum Information and Quantum Optics and
Department of Physics, Fuzhou University, Fuzhou 350116, People\textquoteright s
Republic of China}

\author{Zhen-Biao Yang}

\affiliation{Fujian Key Laboratory of Quantum Information and Quantum Optics and
Department of Physics, Fuzhou University, Fuzhou 350116, People\textquoteright s
Republic of China}

\author{Huaizhi Wu}

\affiliation{Fujian Key Laboratory of Quantum Information and Quantum Optics and
Department of Physics, Fuzhou University, Fuzhou 350116, People\textquoteright s
Republic of China}

\author{Yong Li}

\affiliation{Beijing Computational Science Research Center, Beijing 100193, People\textquoteright s
Republic of China }

\author{Shi-Biao Zheng}

\affiliation{Fujian Key Laboratory of Quantum Information and Quantum Optics and
Department of Physics, Fuzhou University, Fuzhou 350116, People\textquoteright s
Republic of China}
\begin{abstract}
We investigate the dynamics of an optomechanical system where a cavity
with a movable mirror involves a degenerate optical parametric amplifier
and is driven by a periodically modulated laser field. Our results
show that the cooperation between the parametric driving and periodically
modulated cavity driving results in a two-fold squeezing on the movable
cavity mirror that acts as a mechanical oscillator. This allows the
fluctuation of the mechanical oscillator in one quadrature (momentum
or position) to be reduced to a level that cannot be reached by solely
applying either of these two drivings. In addition to the fundamental
interests, e.g., study of quantum effects at the macroscopic level
and exploration of the quantum-to-classical transition, our results
have potential applications in ultrasensitive sensing of force and
motion.
\end{abstract}
\maketitle

\section{Introduction}

Squeezing associated with the mechanical motion of a massive object
\cite{Lecocq_PRX2015,Pirkkalainen_PRL2015,Pontin_PRL2014,Szorkovszky_PRL2013,Schwab_NatPhy2009,Ruskov_prb2005,Rabl_PRL2004,Vanner_PNAS2011,Vanner_NatCom2013}
refers to the reduction of the quantum fluctuation in its position
or momentum below the vacuum level, which is not only important for
fundamental test of quantum theory \cite{Aspelmeyer_PT2012}, such
as exploration of the quantum-classical boundary \cite{Zurek_QCBoundary},
but also have potential applications in high-precision measurement
\cite{HighP_PRD1979,Peano_PRL2015}. In analogy to the standard parametric
techniques applied for squeezing of optical fields, the thermal noise
of a mechanical oscillator can be reduced directly via parametrical
modulation of the mechanical spring constant \cite{Rugar_PRL1991_3db_limit}.
However, even though the mechanical oscillator is initially prepared
in its quantum ground state, the parametric approach failed to generate
a steady-state squeezing of mechanical motion below one half of the
zero-point level (i.e. the well-known 3-dB limit) due to the onset
of instability.

In cavity optomechanical systems \cite{Aspelmeyer_RMP2014,Clerk_njp2008,Zhang_pra2009,Gu_pra2013,Asjad_pra2014,Lv_pra2015,Wang_PRL2013},
theoretical schemes for surpassing the 3-dB limit to realize mechanical
squeezing have been proposed, e.g. by injecting a broad band squeezed
light into the cavity to transfer optical squeezing into mechanical
mode \cite{Jahne_pra2009,Huang_pra2010} or by driving the optical
cavity with two-tone control lasers of different amplitudes combined
with a reservoir engineering technique \cite{Kron_pra2013,Woll_Sci2015},
based on which the experimental demonstration of stationary squeezing
beyond the 3-dB limit was recently achieved \cite{Lei_PRL2016_3db}.
Additionally, it was shown that mechanical squeezing can also be generated
simply by using a periodically amplitude-modulated driving laser \cite{Mari_PRL2009}
or by directly coupling an optical parametric amplifier (OPA) to the
optical cavity \cite{Pinaotey_pra2014} , without the requirement
of classical feedback and of the input of squeezed light \cite{Agarwal_pra2016}.
Despite the advantages of each scheme on certain conditions, it is still highly desirable to further strengthen the mechanical squeezing, and then the following important problems remain open: Does there exist a cooperative effect when the physical processes used
for different methods are applied at the same time? If yes, to what
extent can the mechanical squeezing be enhanced by this cooperative
effect?

\begin{figure}
\includegraphics[width=1\columnwidth]{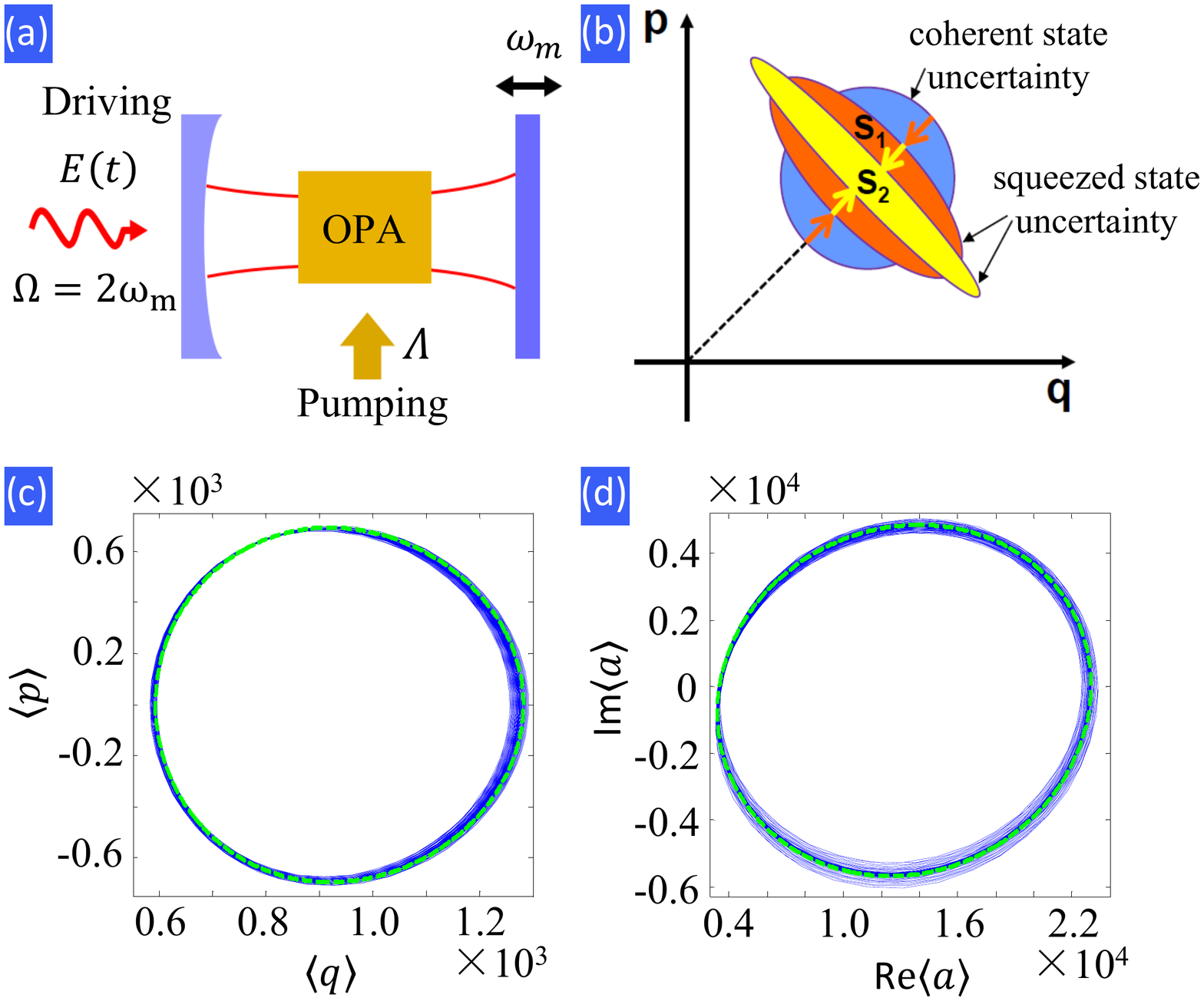}\caption{\label{fig:model}(Color online) (a) Sketch of the optomechanical
setup. The optomechanical cavity that is driven by a periodically
amplitude-modulated laser field contains an OPA, which is pumped by
a laser of the frequency twice of the cavity resonance. See text for
details. (b) Two-fold mechanical squeezing. The fluctuation in one
quadrature of the mechanical mode, reduced to $S_{1}$ by the periodically
amplitude-modulated cavity driving, is further shrunk to $S_{2}$
with the addition of the OPA driving. (c), (d) Phase space trajectories
of the first moments of the mechanical mode for $t=[200\tau,360\tau]$
and optical mode for $t=[20\tau,160\tau]$ with numerical simulations
(blue), and analytical approximations of the asymptotic orbits (green
dash). Parameters are $\left(\kappa,\Delta_{0},g,\gamma_{m}\right)/\omega_{m}=\left(0.1,1.06,4\times10^{-6},10^{-6}\right)$,
$(E_{0},E_{+1},E_{-1})/\omega_{m}=(1.4,0.7,0.7)\times10^{4}$, $\Lambda/\kappa=0.3$,
$\theta=\pi$, $n_{a}=0$, $n_{m}=100$, and $\Omega=2\omega_{m}$.}
\end{figure}

In this paper, we study the quantum dynamics of an optical cavity
that has a movable mirror and contains a degenerate OPA, and which
is driven by a laser field with periodically modulated amplitude,
as shown in Fig. \ref{fig:model}(a). Our results reveal a cooperation-based
enhancement of the squeezing in the fluctuation of the momentum or
position of the cavity mirror. Both the parametric pump driving and
periodically modulated cavity driving contribute to the reduction
of the mechanical fluctuation. The resulting two-fold squeezing exceeds
the squeezing that can be achieved solely by either of these two processes
{[}see Fig. \ref{fig:model}(b){]}. The idea may be generalized to
realize cooperation-based enhancement of other quantum effects in
complex optomechanical systems, e.g., entanglement between two mechanical
oscillators or entanglement between a light field and a mechanical
oscillator \cite{Xuereb_pra2012,Farace_pra2012}.

\section{Theoretical model}

We consider an optomechanical system where a degenerate OPA placed
in a Fabry-Perot cavity of length $L$ and finesse $F$, with one
fixed and partially transmitting mirror, and one movable and totally
reflecting mirror \cite{Agarwal_pra2016,LvXY_PRL2015}. The movable
mirror is treated as a quantum-mechanical harmonic oscillator with
effective mass $m$, frequency $\omega_{m}$, and energy decay rate
$\gamma_{m}$. The cavity mode of resonant frequency $\omega_{c}$
is driven by an external laser of the carrier frequency $\omega_{l}$
(along the cavity axis) with periodically modulated amplitude $E(t)=\sum_{n=-\infty}^{+\infty}E_{n}e^{-in\Omega t}$,
where $\Omega=2\pi/\tau$ with $\tau>0$ being the modulation period,
and the modulation coefficients $\{E_{n}\}$ are related to the power
of the associated sidebands $\{P_{n}\}$ by $\left|E_{n}\right|=\sqrt{2\kappa P_{n}/\hbar\omega_{l}}$,
with $\kappa=\pi c/(2FL)$ being the cavity decay rate due to photon
leakage through the fixed mirror. The degenerate OPA in the optical
cavity is pumped by a coherent field at frequency $2\omega_{p}$,
which leads to the squeezing of cavity field \cite{Walls_book2007,NOTE},
affecting the state of the movable cavity mirror through the optomechanical
coupling. We denote the gain of the OPA by $\Lambda$ (which depends
on the pumping intensity) and the phase of the pump driving as $\theta$.
The total Hamiltonian of the system in the frame rotating at the laser
frequency $\omega_{l}$ can be written as $(\hbar=1)$

\begin{align}
H= & \Delta_{0}a^{\dagger}a+\frac{\omega_{m}}{2}(p^{2}+q^{2})-ga^{\dagger}aq\nonumber \\
 & +i\Lambda(e^{i\theta}a^{\dagger2}e^{-i2\Delta_{p}t}-e^{-i\theta}a^{2}e^{i2\Delta_{p}t})\nonumber \\
 & +i[E(t)a^{\dagger}-E^{*}(t)a].\label{eq:total Hamiltonian}
\end{align}
Here, $\Delta_{0}=\omega_{c}-\omega_{l}$, $\Delta_{p}=\omega_{p}-\omega_{l}$,
$a$ and $a^{\dagger}$ are annihilation and creation operators of
the cavity mode, $q$ and $p$ are the position and momentum operators
for the movable mirror satisfying the standard canonical commutation
relation $[q,\:p]=i$, and $g$=$x_{ZPF}\omega_{c}/L$ is the single-photon
coupling strength between light and mechanical oscillator arising
from the radiation pressure force, with $x_{ZPF}=\sqrt{\hbar/2m\omega_{m}}$
being the zero-point motion of the mechanical mode.

When the mechanical damping and cavity decay are included, the dissipative
dynamics of the open system can be described by the following set
of quantum Langevin equations (QLEs) \cite{Gardiner_book2004}
\begin{align}
\dot{q}= & \omega_{m}p,\nonumber \\
\dot{p}= & -\omega_{m}q-\gamma_{m}p+ga^{\dagger}a+\xi(t),\nonumber \\
\dot{a}= & -(\kappa+i\text{\ensuremath{\Delta}}_{0})a+igaq+E(t)+2\Lambda e^{i\theta}a^{\dagger}e^{-i2\Delta_{p}t}\nonumber \\
 & +\sqrt{2\kappa}a_{in}(t),\label{eq:Langevin}
\end{align}
where both the optical ($a_{in}$) and mechanical ($\xi$) noise operators
have zero-mean value, and the nonzero correlation functions of $a_{in}$
are $\langle a_{in}^{\dagger}(t)a_{in}(t')\rangle=n_{a}\delta(t-t')$
and $\langle a_{in}(t)a_{in}^{\dagger}(t')\rangle=(n_{a}+1)\delta(t-t')$
with $n_{a}=[\exp(\hbar\omega_{c}/k_{B}T)-1]^{-1}$ being the thermal
photon number and that of $\xi(t)$ is given by $\langle{\textstyle \xi(t)\xi(t')\rangle}=\frac{\gamma_{m}}{\omega_{m}}\int\frac{d\omega}{2\pi}e^{-i\omega(t-t')}\omega[1+\mathrm{coth}(\frac{\hbar\omega}{2k_{B}T})]$
\cite{Giovannetti_pra2001,Clerk_RMP2010}. For the specific case where
the mechanical oscillator has a good quality factor $Q\equiv\omega_{m}/\gamma_{m}\gg1$,
$\xi(t)$ becomes delta-correlated $\langle\xi(t)\xi(t')+\xi(t')\xi(t)\rangle/2=\gamma_{m}(2n_{m}+1)\delta(t-t')$
\cite{Benguria_PRL1981,Vitali_pra2007}, which corresponds to the
Markovian process with $n_{m}=[\exp(\hbar\omega_{m}/k_{B}T)-1]^{-1}$
being the mean thermal excitation number in the mechanical mode.

\section{dynamics of the first moments of the optical and mechanical modes}

Suppose that the external drivings are strong enough such that the
intracavity photon number is much larger than 1, we can rewrite each
Heisenberg operator as $O=\left\langle O(t)\right\rangle +\delta O$
($O=q,\,p,\,a$), where $\delta O$ are quantum fluctuation operators
with zero-mean values; and justify that $\left\langle a^{\dagger}(t)a(t)\right\rangle \simeq\left|\left\langle a(t)\right\rangle \right|^{2}$
and $\left\langle a(t)q(t)\right\rangle \simeq\left\langle a(t)\right\rangle \left\langle q(t)\right\rangle $
are valid approximations. Applying the standard linearization techniques
to the QLEs (\ref{eq:Langevin}) and setting $\Delta_{p}=\Omega/2$
for the consideration of mechanical squeezing, we thus obtain the
equations for the first moments of the optical and mechanical modes
\begin{align}
\left\langle \dot{q}(t)\right\rangle = & \omega_{m}\left\langle p(t)\right\rangle ,\nonumber \\
\left\langle \dot{p}(t)\right\rangle = & -\omega_{m}\left\langle q(t)\right\rangle -\gamma_{m}\left\langle p(t)\right\rangle +g\left|\left\langle a(t)\right\rangle \right|^{2},\nonumber \\
\left\langle \dot{a}(t)\right\rangle = & -(\kappa+i\Delta_{0})\left\langle a(t)\right\rangle +ig\left\langle a(t)\right\rangle \left\langle q(t)\right\rangle \nonumber \\
 & +E(t)+2\Lambda e^{i\theta}\left\langle a(t)\right\rangle ^{*}e^{-i\text{\ensuremath{\Omega}}t},\label{eq:mean Langevin}
\end{align}
and the linearized QLEs for the quantum fluctuations
\begin{eqnarray}
\delta\dot{q} & = & \omega_{m}\delta p,\nonumber \\
\delta\dot{p} & = & -\omega_{m}\delta q-\gamma_{m}\delta p+g[\left\langle a(t)\right\rangle \delta a^{\dagger}+\left\langle a(t)\right\rangle ^{*}\delta a]+\xi(t),\nonumber \\
\delta\dot{a} & = & -(\kappa+i\Delta)\delta a+ig\left\langle a(t)\right\rangle \delta q+2\Lambda e^{i\theta}\delta a^{\dagger}e^{-i\Omega t}\nonumber \\
 &  & +\sqrt{2\kappa}a_{in}(t),\label{eq:Linearized Langevin}
\end{eqnarray}
where $\Delta(t)=\Delta_{0}-g\left\langle q(t)\right\rangle $ is
slightly modulated by the mechanical motion.

The phase space trajectories of the first moments $\left\langle O(t)\right\rangle $
can be found by simulating Eq. (\ref{eq:mean Langevin}) for a set
of typical parameters {[}see Figs. \ref{fig:model}(c)-(d){]} \cite{Groblacher_NatPhys2009}.
When the system is far away from the optomechanical instabilities
and multistabilities \cite{Ludwing_njp2008}, the semiclassical dynamics
in the steady state will evolve toward a fixed orbit with a period
being equal to the modulation period of the cavity driving $\tau$.
Moreover, since the two nonlinear terms in Eq. (\ref{eq:mean Langevin})
are both proportional to the coupling strength $g$, the asymptotic
solutions of $\langle O(t)\rangle$ can then be expanded perturbatively
in the powers of $g$ and in terms of the Fourier components for $g\ll\omega_{m}$
\cite{Mari_PRL2009,Mari_njp2012}
\begin{equation}
\begin{array}{ccc}
\left\langle O(t)\right\rangle  & = & \stackrel[j=0]{\infty}{\sum}\stackrel[n=-\infty]{\infty}{\sum}\end{array}O_{n,\,j}e^{in\Omega t}g^{j}.\label{eq:Fourier}
\end{equation}
Substituting Eq. (\ref{eq:Fourier}) into Eq. (\ref{eq:mean Langevin}),
we can then obtain the recursive formulas for the time-independent
coefficients $O_{n,\,j}$ (see Appendix A). By truncating the series
to the first terms with indexes $j=0,\,1,\,...,\,6$ and $n=-1,\,0,\,1$,
we find that the analytical approximations for $\langle O(t)\rangle$
agree well with the numerical results shown in Fig. \ref{fig:model}(c)-(d).
Thus, the linearized dynamics can be evaluated with high accuracy
for the effective optomechanical coupling simply written as
\begin{equation}
G(t)=g_{0}+g_{1}e^{-i\Omega t}+g_{-1}e^{i\Omega t},\label{eq:simple coupling}
\end{equation}
where $g_{n}=|g_{n}|e^{i\phi_{n}}=\frac{1}{\sqrt{2}}\stackrel[j=0]{\infty}{\sum}a_{-n,j}g^{j+1}$
with $n=-1,0,1$. 

\section{quantum fluctuations and Two-fold mechanical squeezing}

To examine the effect of the modulation sidebands ($\sim e^{\pm i\Omega t}$),
we introduce the mechanical annihilation and creation operators $\delta b=(\delta q+i\delta p)/\sqrt{2}$,
$\delta b^{\dagger}=(\delta q-i\delta p)/\sqrt{2}$. Then, the QLEs
for $\delta a$ and $\delta b$ are 
\begin{eqnarray}
\delta\dot{a} & = & -i\Delta\delta a+iG(t)(\delta b^{\dagger}+\delta b)+2\Lambda e^{i\theta}\delta a^{\dagger}e^{-i\Omega t}-\kappa\delta a\nonumber \\
 &  & +\sqrt{2\kappa}a_{in}(t),\nonumber \\
\delta\dot{b} & = & -i\omega_{m}\delta b-\frac{\gamma_{m}}{2}(\delta b-\delta b^{\dagger})+i[G(t)\delta a^{\dagger}+G^{*}(t)\delta a]\nonumber \\
 &  & +\sqrt{\gamma_{m}}b_{in}(t),\label{eq:a=000026b_QLEs}
\end{eqnarray}
with the mechanical noise operator $b_{in}$ satisfying $\text{\ensuremath{\langle}}b_{in}\rangle=0$,
$\langle b_{in}^{\dagger}(t)b_{in}(t')\rangle=n_{m}\delta(t-t'),$
and $\langle b_{in}(t)b_{in}^{\dagger}(t')\rangle=\left(n_{m}+1\right)\delta(t-t').$
We assume that the modulation frequency satisfies $\Omega=2\omega_{m}$
and the carrier frequency of the laser field driving the cavity is
close to the anti-Stokes sideband, which leads to $\Delta=\Delta_{0}-g\langle q(t)\rangle\simeq\omega_{m}$
for weak optomechanical single-photon coupling. We further assume
that the system is working in the resolved sideband regime: $\omega_{m}\gg\kappa$,
and the driving fields are weak: $\omega_{m}\gg|g_{0}|,\,|g_{-1}|,\,|g_{1}|$.
Under these conditions,\textcolor{black}{{} if we substitute the slow
varying fluctuation operators $\delta a(t)=\delta\tilde{a}(t)e^{-i\Delta t}$,
$\delta b(t)=\delta\tilde{b}(t)e^{-i\omega_{m}t}$, $a_{in}(t)=\tilde{a}_{in}(t)e^{-i\Delta t}$
and }$b_{in}(t)=\tilde{b}_{in}(t)e^{-i\omega_{m}t}$\textcolor{black}{{}
into Eq. (\ref{eq:a=000026b_QLEs}), }the terms rotating at $\pm2\omega_{m}$
and $\pm4\omega_{m}$ can be ignored in the rotating wave approximation
(RWA), which leads to
\begin{eqnarray}
\delta\dot{\tilde{a}} & = & ig_{0}\delta\tilde{b}+ig_{1}\delta\tilde{b}^{\dagger}+2\Lambda e^{i\theta}\delta\tilde{a}^{\dagger}-\kappa\delta\tilde{a}+\sqrt{2\kappa}\tilde{a}_{in}(t),\nonumber \\
\delta\dot{\tilde{b}} & = & ig_{0}^{*}\delta\tilde{a}+ig_{1}\delta\tilde{a}^{\dagger}-\frac{\gamma_{m}}{2}\delta\tilde{b}+\sqrt{\gamma_{m}}\tilde{b}_{in}(t).\label{eq:after rotating}
\end{eqnarray}
Note that $\tilde{a}_{in}$ ($\tilde{b}_{in}$) has the same correlation
function as $a_{in}$ ($b_{in}$). We then introduce the optical and
mechanical quadratures with the tilded operators $\delta\tilde{x}=(\delta\tilde{a}+\delta\tilde{a}^{\dagger})/\sqrt{2}$,
$\delta\tilde{y}=(\delta\tilde{a}-\delta\tilde{a}^{\dagger})/i\sqrt{2}$,
$\delta\tilde{q}=(\delta\tilde{b}+\delta\tilde{b}^{\dagger})/\sqrt{2}$,
$\delta\tilde{p}=(\delta\tilde{b}-\delta\tilde{b}^{\dagger})/i\sqrt{2}$,
and the corresponding noise operators $\tilde{x}_{in}=(\tilde{a}_{in}+\tilde{a}_{in}^{\dagger})/\sqrt{2}$,
$\tilde{y}_{in}=(\tilde{a}_{in}-\tilde{a}_{in}^{\dagger})/i\sqrt{2}$,
$\tilde{q}_{in}=(\tilde{b}_{in}+\tilde{b}_{in}^{\dagger})/\sqrt{2}$,
$\tilde{p}_{in}=(\tilde{b}_{in}-\tilde{b}_{in}^{\dagger})/i\sqrt{2}$,
in terms of which the QLEs (\ref{eq:after rotating}) can be rewritten
as

\begin{equation}
\dot{U}t)=\tilde{M}U(t)+N(t),\label{eq:re-written Langevin form}
\end{equation}
where $U(t)=[\delta\tilde{q},\delta\tilde{p},\delta\tilde{x},\delta\tilde{y}]^{T}$,
$N(t)=[\sqrt{\gamma_{m}}\tilde{q}_{in},\sqrt{\gamma_{m}}\tilde{p}_{in},\sqrt{2\kappa}\tilde{x}_{in},\sqrt{2\kappa}\tilde{y}_{in}]$,
and 

\begin{equation}
\tilde{M}=\left[\begin{array}{cccc}
-\frac{\gamma_{m}}{2} & 0 & \mathrm{Im}g_{-} & -\mathrm{Re}g_{-}\\
0 & -\frac{\gamma_{m}}{2} & \mathrm{Re}g_{+} & \mathrm{Im}g_{+}\\
\mathrm{-Im}g_{+} & -\mathrm{Re}g_{-} & -\kappa+2\Lambda\cos\theta & 2\Lambda\sin\theta\\
\mathrm{Re}g_{+} & -\mathrm{Im}g_{-} & 2\Lambda\sin\theta & -\kappa-2\Lambda\cos\theta
\end{array}\right]
\end{equation}
with $g_{\pm}=g_{0}\pm g_{1}$. Note that the stability conditions
derived from the Routh-Hurwitz criterion require the parametric gain
to fulfill $\bar{\Lambda}\equiv2\Lambda/\kappa<1$, the calculation
of which is fussy and will not be shown here.

The mechanical squeezing can be measured by the variance of the tilded
fluctuations $\left\langle \delta\tilde{q}^{2}\right\rangle $ and
$\left\langle \delta\tilde{p}^{2}\right\rangle $, which are just
the first two diagonal elements of the tilded covariance matrix $\tilde{V}_{i,j}(t)=[\langle U_{i}(t)U_{j}(t)\rangle+\langle U_{j}(t)U_{i}(t)\rangle]/2$.
Using Eqs. (\ref{eq:after rotating})-(\ref{eq:re-written Langevin form}),
$\tilde{V}(t)$ in the steady state is dominated by the Lyapunov equation
(see Appendix A)

\begin{equation}
\begin{array}{ccl}
\tilde{M}\tilde{V}+\tilde{V}\tilde{M}^{T} & = & -D\end{array}\label{eq:Lyapunov equation}
\end{equation}
with $D=diag[0,\gamma_{m}(2n_{m}+1),\kappa(2n_{a}+1),\kappa(2n_{a}+1)]$.
Eq. (\ref{eq:Lyapunov equation}) can be analytically solved in the
parameter regime with negligible mechanical damping $\gamma_{m}\approx0$
and null thermal photon number $n_{a}=0$, leading to
\begin{align}
\left\langle \delta\tilde{q}^{2}\right\rangle  & =S_{-}^{\Omega}-\bar{\text{\ensuremath{\Lambda}}}S_{-}^{\Lambda},\;\left\langle \delta\tilde{p}^{2}\right\rangle =S_{+}^{\Omega}+\bar{\text{\ensuremath{\Lambda}}}S_{+}^{\Lambda},\label{eq:covariance_qp}
\end{align}
where $S_{\mp}^{\Omega}=(|g_{0}|^{2}+|g_{1}|^{2}\mp2|g_{0}||g_{1}|\cos\phi_{r})\mathcal{N}^{-1}$,
$S_{\mp}^{\Lambda}=[|g_{0}|^{2}\cos\phi_{r,0}+|g_{1}|^{2}\cos\phi_{r,1}\mp2|g_{0}||g_{1}|\cos(\frac{\phi_{r,0}+\phi_{r,1}}{2})]\mathcal{N}^{-1}$,
$\mathcal{N}=2(1-\bar{\text{\ensuremath{\Lambda}}}^{2})(|g_{0}|^{2}-|g_{1}|^{2}),$
with $\phi_{r}\equiv\phi_{1}-\phi_{0}$, $\phi_{r,0}\equiv\theta-2\phi_{0}$,
and $\phi_{r,1}\equiv\theta-2\phi_{1}$. Eq. (\ref{eq:covariance_qp})
shows that, under the interplay between the periodic cavity driving
and the parametric interaction, the fluctuations of the position and
momentum of the mechanical oscillator strongly depend on the phase
matching condition. To clarify the underlying physics clearly, we
assume $\phi_{r}=\pi,$ $\phi_{r,0}=\pi$ and $\phi_{r,1}=-\pi$,
then the variance of the position and momentum fluctuations reduce
to
\begin{eqnarray}
\left\langle \delta\tilde{q}^{2}\right\rangle  & = & \frac{1}{2}\frac{1+|\frac{g_{1}}{g_{0}}|}{1-|\frac{g_{1}}{g_{0}}|}(1-\bar{\text{\ensuremath{\Lambda}}})^{-1},\\
\left\langle \delta\tilde{p}^{2}\right\rangle  & = & \frac{1}{2}\frac{1-|\frac{g_{1}}{g_{0}}|}{1+|\frac{g_{1}}{g_{0}}|}(1+\bar{\text{\ensuremath{\Lambda}}})^{-1},\label{eq:variance_pp}
\end{eqnarray}
which reveal that the mechanical mode is squeezed in momentum (i.e.
$\left\langle \delta\tilde{p}^{2}\right\rangle <0.5$). Alternatively,
the position squeezing can be achieved by setting $\phi_{r}=0,$ $\phi_{r,0}=0$
and $\phi_{r,1}=0$. More importantly, Eq. (\ref{eq:variance_pp})
shows that the cooperation between the two driving fields results
in a two-fold squeezing: The coefficient ($1-|\frac{g_{1}}{g_{0}}|)/(1+|\frac{g_{1}}{g_{0}}|$)
describes the squeezing effect produced by the periodically modulated
cavity driving, while $(1+\bar{\text{\ensuremath{\Lambda}}})^{-1}$
corresponds to the effect associated with the parametric driving.

\begin{figure}
\includegraphics[width=1\columnwidth]{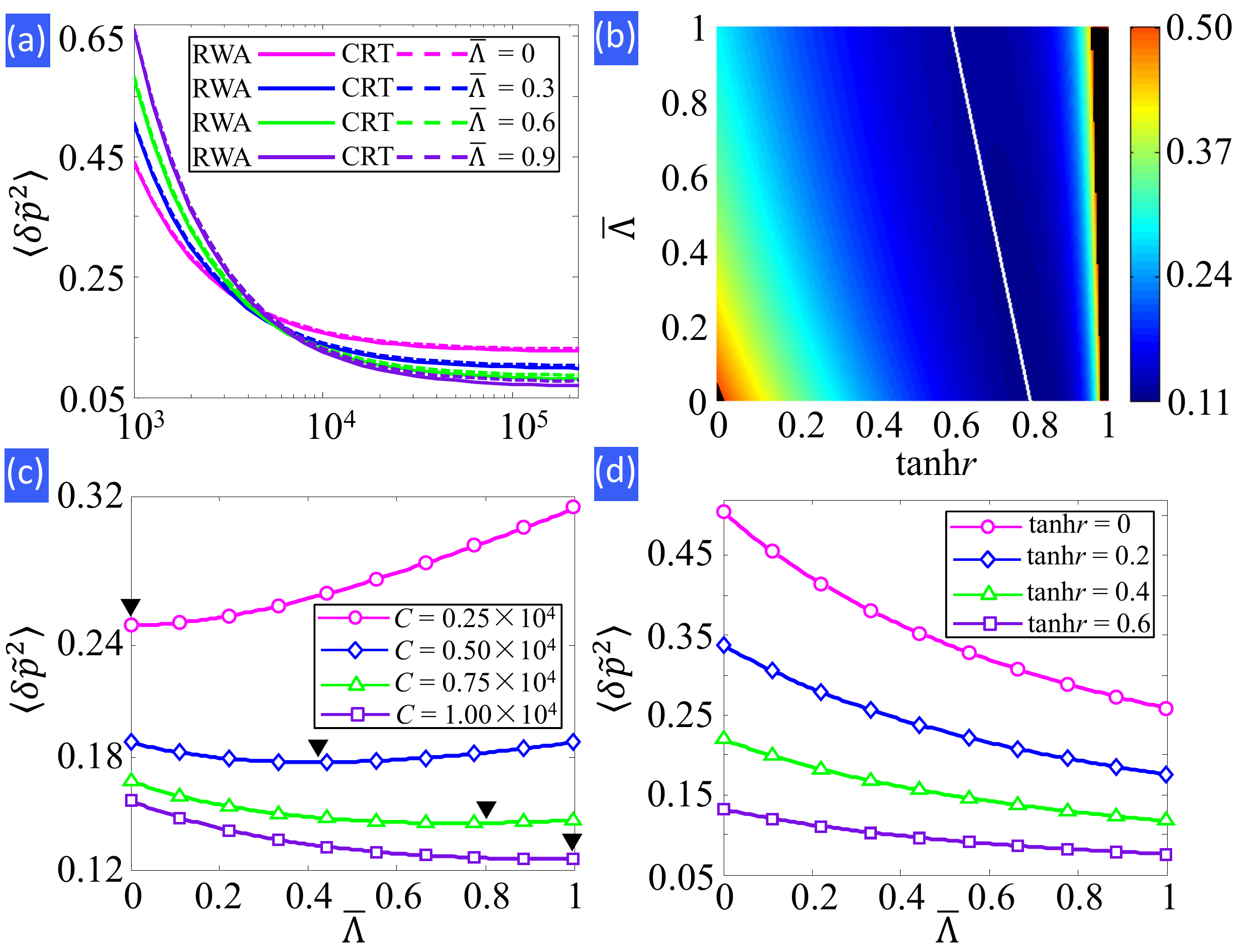}\caption{\label{fig:thr_dimen}(Color online) (a) $\langle\delta\tilde{p}^{2}\rangle$
versus cooperativity parameter $C$ under RWA {[}Eq. (\ref{eq:explicit expression_pp}){]}
and with counter-rotating terms (CRT) included {[}Eq. (\ref{eq:a=000026b_QLEs}){]}
for,\lyxdeleted{Huaizhi Wu}{Mon May  7 18:48:30 2018}{ } $|g_{1}|/|g_{0}|=0.6$,
$|g_{-1}|/|g_{1}|=0.3$, and a set of OPA gains $\bar{\Lambda}$.
(b) $\langle\delta\tilde{p}^{2}\rangle$ versus $\bar{\Lambda}$ and
$\tanh r$ for $C=1.0\times10^{4}$. The black region indicates that
the mechanical oscillator is not squeezed. The white line denotes
the optimal parametric gain $\bar{\Lambda}$ with which the momentum
squeezing reaches its maximum for a given $\tanh r$. (c) $\langle\delta\tilde{p}^{2}\rangle$
versus $\bar{\Lambda}$ for different cooperativity parameters $C$
with $\tanh r=0.6$. The black arrows indicate the optimal squeezing. (d)
$\langle\delta\tilde{p}^{2}\rangle$ versus $\bar{\Lambda}$ for different
modulations of the cavity driving with $C=5\times10^{4}$. The makers
in (c)-(d) indicate the numerical counterpart via the Fourier transformation,
see Appendix C. In all figures other parameters are $\kappa/\omega_{m}=0.1$,
$n_{a}=0$, $n_{m}=100$, $\gamma_{m}/\omega_{m}=10^{-6}$. }
\end{figure}

The two-fold mechanical squeezing can be further understood by introducing
the Bogoliubov mode defined as $\delta B\equiv\delta\tilde{b}\cosh r+e^{i\phi_{r}}\delta\tilde{b}^{\dagger}\sinh r$
with $\tanh r=|g_{1}|/|g_{0}|$ \cite{Pirkkalainen_PRL2015,Kron_pra2013},
which evolves according to the QLEs

\begin{eqnarray}
\delta\dot{B} & = & ig_{B}^{*}\delta\tilde{a}-\frac{\gamma_{m}}{2}\delta B+\sqrt{\gamma_{m}}B_{in},\nonumber \\
\delta\dot{\tilde{a}} & = & -\kappa\delta\tilde{a}+ig_{B}\delta B+2\text{\ensuremath{\Lambda}}e^{i\theta}\delta\tilde{a}^{\dagger}+\sqrt{2\kappa}\tilde{a}_{in},\label{eq:Bogliubov}
\end{eqnarray}
with $g_{B}=\sqrt{|g_{0}|^{2}-|g_{1}|^{2}}e^{i\phi_{0}}$. Since the
vacuum state of the Bogoliubov mode corresponds to a squeezed state,
the noise input $B_{in}$ for $\delta B$ has zero mean and the nonzero
correlation functions $\langle B_{in}^{\dagger}(t)B_{in}(t')\rangle=[(n_{m}+1)\sinh^{2}r+n_{m}\cosh^{2}r]\delta(t-t')$
and $\langle B_{in}(t)B_{in}^{\dagger}(t')\rangle=[(n_{m}+1)\cosh^{2}r+n_{m}\sinh^{2}r]\delta(t-t').$
By applying the adiabatic approximation for $\kappa\gg|g_{B}|$ (i.e.
$\delta\dot{\tilde{a}}$=0) \cite{Agarwal_pra2016}, and considering
the phase matching condition ($\phi_{r,0}=\pi$) for momentum squeezing,
we find that the variance of the quadrature $\delta p_{B}=\frac{1}{\sqrt{2}i}(\delta B-\delta B^{\dagger})$
in the steady state reads (see Appendix B)

\begin{eqnarray}
\langle\delta p_{B}^{2} & \rangle= & \frac{\kappa\gamma_{m}(1+\bar{\text{\ensuremath{\Lambda}}})}{4\left|g_{B}\right|^{2}}(2\sinh^{2}r+1)(2n_{m}+1)\nonumber \\
 &  & +\frac{1}{2(1+\bar{\text{\ensuremath{\Lambda}}})}(2n_{a}+1).
\end{eqnarray}
For $\phi_{r}=\pi$, the variance of the momentum fluctuation for
the original mechanical mode has a simply analytical form

\begin{eqnarray}
\langle\delta\tilde{p}^{2}\rangle & = & (\cosh r-\sinh r)^{2}\langle\delta p_{B}^{2}\rangle,
\end{eqnarray}
which is exactly the result of Eq. (\ref{eq:variance_pp}) for $\gamma_{m}=0$,
$n_{a}=0$. This result can be roughly explained as follows: The periodically
modulated cavity driving produces a squeezing effect on the momentum
fluctuation of the mechanical mode, which mathematically corresponds
to converting the normal mechanical mode into the Bogoliubov mode
through a unitary transformation equivalent to a squeezed operator.
As a consequence, the ``momentum'' fluctuation of the Bogoliubov
mode at the ``vacuum'' level corresponds to the normal momentum
fluctuation below the vacuum level ($\left\langle \delta\tilde{p}^{2}\right\rangle <0.5$).
The parametric driving further reduces the ``momentum'' fluctuation
of the Bogoliubov mode below the ``vacuum'' level, resulting in
a second squeezing effect.

\section{The effect of mechanical damping and experimental feasibility}

Considering the effect of the mechanical damping $\gamma_{m}\neq0$,
the variances of the fluctuations $\left\langle \delta\tilde{q}^{2}\right\rangle $
and $\left\langle \delta\tilde{p}^{2}\right\rangle $ can again be
calculated by the Lyapunov equation (\ref{eq:Lyapunov equation}).
As an example, when the phases $\phi_{0}=0$, $\phi_{r}=\pi$ and
$\theta=\pi$ are set and the cooperativity parameter $C=4|g_{0}|^{2}/(\kappa\gamma_{m})$
is large so that $\tilde{C}\equiv C(1-\tanh^{2}r)\gg2(1+\bar{\Lambda})$,
the variance of the momentum is approximately given by 

\begin{eqnarray}
\left\langle \delta\tilde{p}^{2}\right\rangle  & \approx & \frac{(\cosh r-\sinh r)^{2}}{2(1+\bar{\Lambda})}\nonumber \\
 &  & +(2n_{m}+1)[\frac{1+\bar{\Lambda}}{\tilde{C}}+\frac{\gamma_{m}}{4\kappa(1+\bar{\Lambda})}].\label{eq:explicit expression_pp}
\end{eqnarray}
which agrees well with its numerical counterpart obtained by simulation
of Eq. (\ref{eq:a=000026b_QLEs}), as shown in Fig. \ref{fig:thr_dimen}(a).
For $|g_{1}|\simeq|g_{0}|$ (corresponding to $\tanh r\rightarrow1$),
the effective coupling between the Bogoliubov mode and the cavity
mode becomes negligible, and $\left\langle \delta\tilde{p}^{2}\right\rangle \simeq n_{m}+\frac{1}{2}$
is mainly determined by the thermal occupation of the mechanical mode
with $\gamma_{m}\ll\kappa$, implying that the mechanical mode is
not squeezed {[}see Fig. \ref{fig:thr_dimen}(b){]} \cite{Woll_Sci2015}.
In this case, the self-cooling of the mechanical oscillator through
the photon-phonon sideband coupling is suppressed \cite{Gigan_Nature2006,Kleckner_Nature2006,Marquardt_PRL2007},
therefore, the mechanical oscillator may stay far away from the ground
state \cite{Kron_pra2013,Schonburg_pra2016}. Generally, there exists
an optimal squeezing for $\langle\delta\tilde{p}^{2}\rangle$ corresponding
to the best efficiency of the cooperation between the two driving
fields, which can be readily found by setting $\textrm{d}\langle\delta\tilde{p}^{2}\rangle/\textrm{d}\bar{\Lambda}=0$
for an appropriate amplitude modulation $|g_{1}|/|g_{0}|$ (i.e. a
given $\tanh r$). For $\tilde{C}\gg1$, $\left\langle \delta\tilde{p}^{2}\right\rangle $
reaches its minimum when the optimal parametric gain satisfies $\bar{\Lambda}_{opt}=\frac{\text{\ensuremath{\eta}}}{2}\text{(}1+\sqrt{1+\frac{\tilde{C}}{\eta}})-1$
with $\eta=\frac{(\cosh r-\sinh r)^{2}}{n_{m}+\frac{1}{2}}+\frac{\gamma_{m}}{\kappa},$
which is indicated in Fig. \ref{fig:thr_dimen}(b). Note that an effective
cooperation implies a non-negative $\bar{\Lambda}_{opt}$, which imposes
a threshold of $\tilde{C}_{thr}=4(\eta^{-1}-1)$ on $\tilde{C}$,
namely $\tilde{C}>\tilde{C}_{thr}$. In addition, the stability condition
$\bar{\Lambda}_{opt}<1$ requires $\tilde{C}<\tilde{C}_{ins}$ with
$\tilde{C}_{ins}=8(2\eta^{-1}-1)$, beyond which the best cooperation
efficiency always appears at $\bar{\Lambda}_{opt}\rightarrow1$, in
vicinity of instability, see Fig. \ref{fig:thr_dimen}(c) for the
example of $\tanh r=0.6$, where we find $\tilde{C}_{thr}\approx1.6\times10^{3}$
($C_{thr}\approx2.5\times10^{3}$) and $\tilde{C}_{ins}\approx6.4\times10^{3}$
($C_{ins}\approx10^{4}$).

Considering the set of experimentally feasible parameters \cite{Groblacher_NatPhys2009}:
$L=25$ mm, $F=1.4\times10^{4}$, $\omega_{m}/2\pi=1$ MHz, $Q=10^{6}$,
$m=150$ ng, $T=5$ mK and the power of the carrier component $P_{0}=1$
mW of the driving laser ($\lambda=1064$ nm), we show in Fig. \ref{fig:thr_dimen}(d)
that the degree of squeezing for the mechanical momentum with $C=5\times10^{4}$
and thermal occupation $n_{m}=100$ is monotonically improved as the
dimensionless parametric gain $\bar{\Lambda}$ increases for $\tanh r=0$,
$0.2$, $0.4$, $0.6$ due to $\tilde{C}>\tilde{C}_{ins}$. Under
this condition, the momentum fluctuation is reduced from 0.219 (3.57
dB) to 0.117 (6.29 dB) for $\tanh r=0.4$, and from 0.132 (5.79 dB)
to 0.0756 (8.21 dB) for $\tanh r=0.6$ as the parametric gain $\bar{\Lambda}$
is increased from 0 to the optimal value 0.99. The momentum squeezing
can be further increased for a larger $C/n_{m}$ ratio (the so-called
quantum cooperativity) under the best efficiency of the two-field
cooperation. Our results clearly show that, with suitable choice of
the system parameters, both the cavity driving and parametric interaction
significantly contribute to the reduction of the mechanical momentum
fluctuation; their cooperation is important for realization of a strong
mechanical squeezing.

\section{Conclusion}

In summary, we have shown that the parametric driving and the periodically
modulated cavity driving, simultaneously applied to a cavity optomechanical
system, can result in a two-fold squeezing effects on the mechanical
oscillator. This enables implementation of strong squeezing for a
macroscopic oscillator, which exceeds the result that is solely produced
by either of these two drivings. Our results show that different physical
processes, each producing a weak quantum effect, can cooperate to
enhance the quantum effect. Our idea can be generalized to more complex
optomechanical systems to realize two-fold two-mode squeezing, offering
a possibility to produce strong mechanical-mechanical or optomechanical
entanglement that can exceed the bound imposed by present methods.
\begin{acknowledgments}
This work was supported by the National Natural Science Foundation
of China under Grants No. 11774058 and No. 11774024, and the Natural
Science Foundation of Fujian Province under Grant No. 2017J01401.
\end{acknowledgments}

\appendix

\section{Periodic motion and quantum dynamics of the mechanical oscillator
in the steady state}

The time-independent coefficients $O_{n,\,j}$ in the Fourier expansion
of $\left\langle O(t)\right\rangle $ ($O=q,p,a$) given by Eq. (\ref{eq:Fourier})
can be found by substituting Eq. (\ref{eq:Fourier}) into Eq. (\ref{eq:mean Langevin}),
leading to the following recursive formulas
\begin{eqnarray}
 &  & p_{n,0}=0,\;q_{n,0}=0,\label{Fourier_zeroth}\\
a_{n,0} & = & \frac{\left[\kappa-i\left(\Delta_{0}-(n+1)\Omega\right)\right]E_{-n}+2\Lambda e^{i\theta}E_{n+1}}{\left[\kappa-i\left(\Delta_{0}-(n+1)\Omega\right)\right]\left[\kappa+i\left(\Delta_{0}+n\Omega\right)\right]-4\Lambda^{2}},\nonumber 
\end{eqnarray}
corresponding to the zeroth-order perturbation with respect to $g$,
and with $j>0$,
\begin{eqnarray}
q_{n,j} & = & \omega_{m}\stackrel[k=0]{j-1}{\sum}\stackrel[m=-\infty]{\infty}{\sum}\frac{a_{m,k}^{*}a_{n+m,j-k-1}}{\omega_{m}^{2}-(n\Omega)^{2}+i\gamma_{m}n\Omega},\nonumber \\
p_{n,j} & = & \frac{in\Omega}{\omega_{m}}q_{n,j},\label{Fourier_nth}\\
a_{n,j} & = & \frac{2\Lambda e^{i\theta}a_{-n-1,j}^{*}}{\kappa+i(\Delta_{0}+n\Omega)}+i\stackrel[k=0]{j-1}{\sum}\stackrel[m=-\infty]{\infty}{\sum}\frac{a_{m,k}q_{n-m,j-k-1}}{\kappa+i(\Delta_{0}+n\Omega)}.\nonumber 
\end{eqnarray}
Using Eq. (\ref{Fourier_zeroth}) and (\ref{Fourier_nth}), the quantum
dynamics of the mechanical oscillator can be studied through the linearized
QLEs (\ref{eq:Linearized Langevin}). 

We introduce the amplitude and phase quadratures of the cavity mode
as $\delta x=(\delta a+\delta a^{\dagger})/\sqrt{2}$, $\delta y=(\delta a-\delta a^{\dagger})/i\sqrt{2}$
and the analogous input quantum noise quadratures as $\delta x_{in}=(\delta a_{in}+\delta a_{in}^{\dagger})/\sqrt{2}$,
$\delta y_{in}=(\delta a_{in}-\delta a_{in}^{\dagger})/i\sqrt{2}$
for convenience. Then the time-dependent equations of motion for the
quantum fluctuations $u(t)=[\delta q,\delta p,\delta x,\delta y]^{T}$
arise as 

\begin{equation}
\dot{u}(t)=M(t)u(t)+n(t),\label{eq:re-writev Langevin}
\end{equation}
with the drift matrix
\[
M(t)=\left[\begin{array}{cccc}
0 & \omega_{m} & 0 & 0\\
-\omega_{m} & -\gamma_{m} & 2G_{x}(t) & 2G_{y}(t)\\
-2G_{y}(t) & 0 & -\kappa+2\Lambda\cos\bar{\theta} & \Delta-2\Lambda\sin\bar{\theta}\\
2G_{x}(t) & 0 & -\Delta-2\Lambda\sin\bar{\theta} & -\kappa-2\Lambda\cos\bar{\theta}
\end{array}\right],
\]
and the diffusion $n(t)=[0,\xi(t),\sqrt{2\kappa}\delta x_{in},\sqrt{2\kappa}\delta y_{in}]^{T}$
being the noise sources. Here $\bar{\theta}=\Omega t-\theta$, and
$G_{x}$, $G_{y}$ are real part and imaginary part of the effective
optomechanical coupling $G(t)\equiv g\left\langle a(t)\right\rangle /\sqrt{2}$.
If all the eigenvalues of the matrix $M(t)$ have negative real parts
at any time (i.e. the Routh-Hurwitz criterion) \cite{pra1987_Dejesus},
the system will be in stable in the steady state. On the other hand,
since the system in the steady state will evolve into an asymptotic
Gaussian state for a Gaussian-typed of noise \cite{RMP2012_Weed},
we can then characterize the second moments of the quadratures of
the asymptotic state through the covariance matrix (CM) $V(t)$, with
the matrix elements being
\begin{equation}
\begin{array}{ccc}
V_{k,l}(t) & = & \langle u_{k}(t)u_{l}^{\dagger}(t)+u_{l}^{\dagger}(t)u_{k}(t)\rangle/2.\end{array}\label{eq:Covariance matrix}
\end{equation}
From Eqs. (\ref{eq:re-writev Langevin}) and (\ref{eq:Covariance matrix}),
we can easily derive a linear differential equation governing the
evolution of the CM $V(t)$
\begin{equation}
\begin{array}{ccc}
\dot{V}(t) & = & M(t)V(t)+V(t)M^{T}(t)+D,\end{array}\label{eq:differential equation}
\end{equation}
where $M(t)^{T}$ is the transpose matrix of $M(t)$, and $\begin{array}{ccc}
D & = & diag[0,\gamma_{m}(2n_{m}+1),\kappa(2n_{a}+1),\kappa(2n_{a}+1)]\end{array}$ is a diagonal noise correlations matrix, defined by $\delta(t-t')D_{k,l}=\langle n_{k}(t)n_{l}^{\dagger}(t')+n_{l}^{\dagger}(t')n_{k}(t)\rangle/2$.
The first two diagonal elements $V_{11}(t)=\left\langle \delta q(t)^{2}\right\rangle $,
$V_{22}(t)=\left\langle \delta p(t)^{2}\right\rangle $ of $V(t)$
represent the variances of the fluctuations in the mechanical position
and momentum, and the last two terms $V_{33}(t)=\left\langle \delta x(t)^{2}\right\rangle $,
$V_{44}(t)=\left\langle \delta y(t)^{2}\right\rangle $ represent
the variances of the fluctuations in the amplitude and phase of the
cavity mode. The mechanical oscillator is position- or momentum-squeezed
if either $\left\langle \delta q(t)^{2}\right\rangle <1/2$ or $\left\langle \delta p(t)^{2}\right\rangle <1/2$
in the steady state. The degree of the squeezing can be expressed
in the dB unit, which can be calculated by $-10\log_{10}\frac{\left\langle \delta o(t)^{2}\right\rangle }{\left\langle \delta o(t)^{2}\right\rangle _{vac}}$
(or $o=p,\,q$), with $\left\langle \delta q(t)^{2}\right\rangle _{vac}=\left\langle \delta p(t)^{2}\right\rangle _{vac}=1/2$
being the position and momentum variances of the vacuum state.

\begin{figure}
\includegraphics[width=0.85\columnwidth]{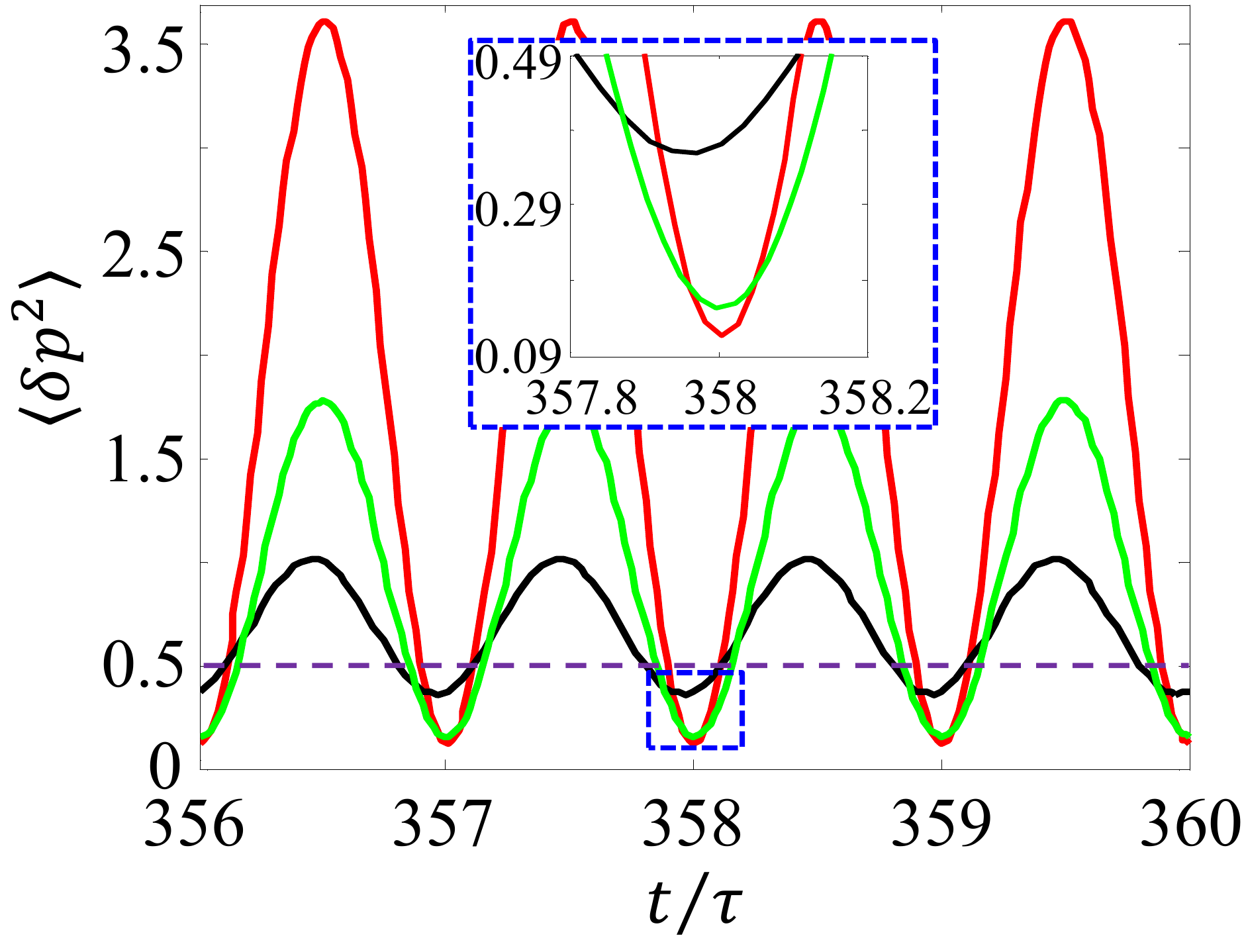}\caption{\label{fig;pp_t_three}(Color online) Variances of the mechanical
momentum fluctuation $\left\langle \delta p(t)^{2}\right\rangle $
versus rescaled time $t/\tau$ for the modulation sidebands driving
amplitudes $E_{\pm1}$ and the parametric gain $\Lambda$ being (i)
$E_{\pm1}=0$, $\text{\ensuremath{\Lambda}/\ensuremath{\kappa}=0.3}$
(black), (ii) $E_{\pm1}/\omega_{m}=0.7\times10^{4}$, $\Lambda/\kappa=0$
(green), (iii) $E_{\pm1}/\omega_{m}=0.7\times10^{4}$, $\text{\ensuremath{\Lambda}/\ensuremath{\kappa}=0.3}$
(red). The dash line represents the momentum variance of the vacuum
state $\left\langle \delta p(t)^{2}\right\rangle _{vac}=0.5$. Other
parameters are the same as in Fig. 1(c),(d).}
\end{figure}

Recalling that the asymptotic behavior of the first moments of the
mechanical mode and the cavity mode is $\tau=2\pi/\Omega$ periodic
in the steady state, then we can find that the drift matrix $M(t)$,
which is related to $\langle a(t)\rangle$ and $\langle q(t)\rangle$,
satisfies $M(t+\tau)=M(t)$ and therefore $V(t+\tau)=V(t)$ according
to the Floquet theory \cite{www_Teschl}. By solving the evolutional
equation (\ref{eq:differential equation}) of the CM $V(t)$, we have
calculated the time-dependent variances of the mechanical momentum
$\left\langle \delta p(t)^{2}\right\rangle $ for the optomechanical
system with (i) OPA, (ii) periodic driving, and (iii) both OPA and
periodic driving, as shown in Fig.\ref{fig;pp_t_three}. It has been
realized that the cavity solely pumped by parametric interaction (with
$E_{\pm1}=0$) \cite{Agarwal_pra2016} and solely modulated by periodic
driving ($\Lambda=0$) \cite{Mari_PRL2009} can both lead to mechanical
squeezing, the degree of which (corresponding to the minimum of $\left\langle \delta p(t)^{2}\right\rangle $)
can reach 1.44 dB ($\left\langle \delta p(t)^{2}\right\rangle _{min}=0.359$)
for $\Lambda/\kappa=0.3$, and 5.13 dB ($\left\langle \delta p(t)^{2}\right\rangle _{min}=0.153$)
for $E_{\pm1}/\omega_{m}=0.7\times10^{4}$, respectively. However,
we note that, by combining OPA and periodic driving simultaneously,
the mechanical squeezing will be greatly enhanced, the degree of momentum
squeezing can achieve as large as 6.31 dB ($\left\langle \delta p(t)^{2}\right\rangle _{min}=0.117$),
which is far beyond the 3 dB limit, required for ultrahigh-precision
measurements. 

\section{The steady-state ``momentum'' fluctuation of the Bogoliubov mode }

The QLEs (\ref{eq:Bogliubov}) can be solved in the adiabatic approximation
under the condition of $\kappa\gg|g_{B}|$. For this purpose, we rewrite
the equations of motion for $\delta\tilde{a}$ and $\delta\tilde{a}^{\dagger}$:

\begin{eqnarray}
\delta\dot{\tilde{a}} & = & -\kappa\delta\tilde{a}+ig_{B}\delta B+2\text{\ensuremath{\Lambda}}e^{i\theta}\delta\tilde{a}^{\dagger}+\sqrt{2\kappa}\tilde{a}_{in},\nonumber \\
\delta\dot{\tilde{a}}^{\dagger} & = & -\kappa\delta\tilde{a}^{\dagger}-ig_{B}^{*}\delta B^{\dagger}+2\text{\ensuremath{\Lambda}}e^{-i\theta}\delta\tilde{a}+\sqrt{2\kappa}\tilde{a}_{in}^{\dagger}.
\end{eqnarray}
Setting $\delta\dot{\tilde{a}}=\delta\dot{\tilde{a}}^{\dagger}=0$
and $\theta-2\phi_{0}=0$, it is readily to find

\begin{eqnarray}
\delta\tilde{a} & = & i\frac{g_{B}}{\kappa(1-\bar{\text{\ensuremath{\Lambda}}}^{2})}(\delta B+\text{\ensuremath{\bar{\text{\ensuremath{\Lambda}}}}}\delta B^{\dagger})\nonumber \\
 &  & +\frac{\sqrt{2\kappa}}{\kappa(1-\bar{\text{\ensuremath{\Lambda}}}^{2})}(\tilde{a}_{in}+\text{\ensuremath{\bar{\text{\ensuremath{\Lambda}}}}}e^{i\theta}\tilde{a}_{in}^{\dagger}).\label{Adiabatic_approxi}
\end{eqnarray}
Inserting Eq. (\ref{Adiabatic_approxi}) into $\delta\dot{B}=ig_{B}^{*}\delta\tilde{a}-\frac{\gamma_{m}}{2}\delta B+\sqrt{\gamma_{m}}B_{in},$
we have

\begin{eqnarray}
\delta\dot{B} & = & -\frac{|g_{B}|^{2}}{\kappa(1-\bar{\text{\ensuremath{\Lambda}}}^{2})}(\delta B+\bar{\text{\ensuremath{\Lambda}}}\delta B^{\dagger})+\sqrt{\gamma_{m}}B_{in}\nonumber \\
 &  & +\frac{ig_{B}^{*}\sqrt{2\kappa}}{\kappa(1-\bar{\text{\ensuremath{\Lambda}}}^{2})}(\delta\tilde{a}_{in}+\bar{\text{\ensuremath{\Lambda}}}e^{i\theta}\delta\tilde{a}_{in}^{\dagger}).\label{eq:squeezing effect}
\end{eqnarray}
Here, the term $\gamma_{m}$ in the coefficient of $\delta B$ is
safely ignored in our parameter regime. Then the equation of motion
for the ``momentum'' of the Bogoliubov mode $\delta p_{B}$ is given
by

\begin{eqnarray}
\delta\dot{p}_{B} & = & -\frac{|g_{B}|^{2}}{\kappa(1+\bar{\text{\ensuremath{\Lambda}}})}\delta p_{B}+d(t)+v(t),\label{eq:Momentum operator}
\end{eqnarray}
where $d(t)=\frac{1}{i}\sqrt{\frac{\gamma_{m}}{2}}(B_{in}-B{}_{in}^{\dagger})$
and $v(t)=\frac{g_{B}^{*}\sqrt{\kappa}}{\kappa(1+\bar{\text{\ensuremath{\Lambda}}})}(\tilde{a}_{in}-\tilde{a}_{in}^{\dagger})e^{i\theta}$,
whose correlation functions are $\langle d(t)d(t')\rangle=\frac{\gamma_{m}}{2}(2\sinh^{2}r+1)(2n_{m}+1)\delta(t-t')$
and $\langle v(t)v(t')\rangle=\frac{|g_{B}|^{2}}{\kappa(1+\bar{\text{\ensuremath{\Lambda}}})^{2}}(2n_{a}+1)\delta(t-t')$,
respectively. Based on these equations, we obtain the equation for
$\left\langle \delta p_{B}^{2}\right\rangle $ as 

\begin{eqnarray}
\frac{\partial\langle\delta p_{B}^{2}\rangle}{\partial t} & = & -\frac{2|g_{B}|^{2}}{\kappa(1+\bar{\text{\ensuremath{\Lambda}}})}\langle\delta p_{B}^{2}\rangle+\frac{|g_{B}|^{2}}{\kappa(1+\bar{\text{\ensuremath{\Lambda}}})^{2}}(2n_{a}+1)\nonumber \\
 &  & +\frac{\gamma_{m}}{2}(2\sinh^{2}r+1)(2n_{m}+1).
\end{eqnarray}
As a result, the steady-state ``momentum'' fluctuation of the Bogoliubov
mode can be solved by setting $\frac{\partial\langle\delta p_{B}^{2}\rangle}{\partial t}=0$,
giving rise to Eq. (16) in the main text.

\section{Optomechanical squeezing in the rotating wave approximation }

Taking the Fourier transform of Eq. (\ref{eq:re-written Langevin form})
by using $f(t)=\frac{1}{2\pi}\intop_{-\infty}^{+\infty}f(\omega)e^{-i\omega t}d\omega$
and $f^{\dagger}(t)=\frac{1}{2\pi}\int_{-\infty}^{+\infty}f^{\dagger}(-\omega)e^{-i\omega t}d\omega$,
we obtain the position and momentum fluctuations of the movable mirror
in the frequency domain, i.e.

\begin{equation}
\begin{array}{ccc}
\delta\tilde{q}(\omega) & = & A_{1}(\omega)\tilde{x}_{in}(\omega)+B_{1}(\omega)\tilde{y}_{in}\\
 &  & +E_{1}(\omega)\tilde{q}_{in}(\omega)+F_{1}(\omega)\tilde{p}_{in},\\
\delta\tilde{p}(\omega) & = & A_{2}(\omega)\tilde{x}_{in}(\omega)+B_{2}(\omega)\tilde{y}_{in}\\
 &  & +E_{2}(\omega)\tilde{q}_{in}(\omega)+F_{2}(\omega)\tilde{p}_{in},
\end{array}\label{eq:Fourier transform form}
\end{equation}
where

\begin{equation}
\begin{array}{ccl}
A_{1}(\omega) & = & -\frac{\sqrt{2\kappa}i}{d(\omega)}\{[\Lambda(\alpha_{0}-\alpha_{1})+iu(\omega)\mathrm{Im}(g_{0}-g_{1})]v(\omega)\\
 &  & +i(\left|g_{0}\right|^{2}-\left|g_{1}\right|^{2})\mathrm{Im}(g_{0}-g_{1})\},\\
B_{1}(\omega) & = & \frac{\sqrt{2\kappa}}{d(\omega)}\{[\Lambda(\beta_{0}-\beta_{1})-u(\omega)\mathrm{Re}(g_{0}-g_{1})]v(\omega)\\
 &  & -(\left|g_{0}\right|^{2}-\left|g_{1}\right|^{2})\mathrm{Re}(g_{0}-g_{1})\},\\
E_{1}(\omega) & = & \frac{\sqrt{\gamma_{m}}}{d(\omega)}\{[u(\omega)^{2}-4\Lambda^{2}]v(\omega)\\
 &  & +(\left|g_{0}\right|^{2}-\left|g_{1}\right|^{2})u(\omega)+\Lambda(\Gamma_{0}-\Gamma_{1})\},\\
F_{1}(\omega) & = & \frac{\sqrt{\gamma_{m}}}{d(\omega)}i\Lambda[(g_{0}-g_{1})^{2}e^{-i\theta}-(g_{0}^{*}-g_{1}^{*})^{2}e^{i\theta}],\\
A_{2}(\omega) & = & \frac{\sqrt{2\kappa}}{d(\omega)}\{[\Lambda(\beta_{0}+\beta_{1})+u(\omega)\mathrm{Re}(g_{0}+g_{1})]v(\omega)\\
 &  & +(\left|g_{0}\right|^{2}-\left|g_{1}\right|^{2})\mathrm{Re}(g_{0}+g_{1})\},\\
B_{2}(\omega) & = & \frac{\sqrt{2\kappa}i}{d(\omega)}\{[\Lambda(\alpha_{0}+\alpha_{1})-iu(\omega)\mathrm{Im}(g_{0}+g_{1})]v(\omega)\\
 &  & -i(\left|g_{0}\right|^{2}-\left|g_{1}\right|^{2})\mathrm{Im}(g_{0}+g_{1})\},\\
E_{2}(\omega) & = & \frac{\sqrt{\gamma_{m}}}{d(\omega)}i\Lambda[(g_{0}+g_{1})^{2}e^{-i\theta}-(g_{0}^{*}+g_{1}^{*})^{2}e^{i\theta}],\\
F_{2}(\omega) & = & \frac{\sqrt{\gamma_{m}}}{d(\omega)}\{[u(\omega)^{2}-4\Lambda^{2}]v(\omega)\\
 &  & +(\left|g_{0}\right|^{2}-\left|g_{1}\right|^{2})u(\omega)-\Lambda(\Gamma_{0}-\Gamma_{1})\},
\end{array}
\end{equation}
with $\alpha_{0}=g_{0}e^{-i\theta}-g_{0}^{*}e^{i\theta}$, $\alpha_{1}=g_{1}e^{-i\theta}-g_{1}^{*}e^{i\theta}$
, $\beta_{0}=g_{0}e^{-i\theta}+g_{0}^{*}e^{i\theta}$, $\beta_{1}=g_{1}e^{-i\theta}+g_{1}^{*}e^{i\theta}$
, $\Gamma_{0}=g_{0}^{2}e^{-i\theta}+g_{0}^{*2}e^{i\theta}$, $\Gamma_{1}=g_{1}^{2}e^{-i\theta}+g_{1}^{*2}e^{i\theta}$,
$u(\omega)=\frac{\gamma_{m}}{2}-i\omega$, $v(\omega)=\kappa-i\omega$
and

\begin{equation}
\begin{array}{ccl}
d(\omega) & = & [u(\omega)v(\omega)+(\left|g_{0}\right|^{2}-\left|g_{1}\right|^{2})]^{2}-4\Lambda^{2}v(\omega)^{2}.\end{array}
\end{equation}
The first two terms in $\delta\tilde{q}(\omega)$ and $\delta\tilde{p}(\omega)$
originate from the radiation pressure contribution, and the last two
terms are from the thermal noise contribution. Without optomechanical
coupling ($g_{0}=g_{1}=0$), the mechanical mode subjected to the
purely thermal noise will make quantum Brownian motion leading to
$\delta\tilde{q}(\omega)=$$\frac{\sqrt{\gamma_{m}}}{\frac{\gamma_{m}}{2}-i\omega}\tilde{q}_{in}$
and $\delta\tilde{p}(\omega)=$$\frac{\sqrt{\gamma_{m}}}{\frac{\gamma_{m}}{2}-i\omega}\tilde{p}_{in}$.
The expressions of the spectra for the position and momentum fluctuations
of the mechanical mode are ($Z=\tilde{q},\,\tilde{p}$)

\begin{equation}
\begin{array}{ccl}
S_{Z}(\omega) & = & \frac{1}{4\pi}\int_{-\infty}^{+\infty}d\omega'e^{-i(\omega'+\omega)t}\\
 &  & \times[\left\langle \delta Z(\omega)\delta Z(\omega')\right\rangle +\left\langle \delta Z(\omega')\delta Z(\omega)\right\rangle ],
\end{array}\label{eq:power spectrum}
\end{equation}
which can be solved by using the correlation functions of the noise
sources in the frequency domain \cite{Agarwal_pra2016}

\begin{equation}
\begin{array}{cll}
\left\langle \tilde{q}_{in}(\omega)\tilde{q}_{in}(\omega')\right\rangle  & = & \left\langle \tilde{p}_{in}(\omega)\tilde{p}_{in}(\omega')\right\rangle \\
 & = & (n_{m}+\frac{1}{2})2\pi\delta(\omega+\omega'),\\
\left\langle \tilde{q}_{in}(\omega)\tilde{p}_{in}(\omega')\right\rangle  & = & -\left\langle \tilde{p}_{in}(\omega)\tilde{q}_{in}(\omega')\right\rangle \\
 & = & \frac{i}{2}2\pi\delta(\omega+\omega'),\\
\left\langle \tilde{x}_{in}(\omega)\tilde{x}_{in}(\omega')\right\rangle  & = & \left\langle \tilde{y}_{in}(\omega)\tilde{y}_{in}(\omega')\right\rangle \\
 & = & (n_{a}+\frac{1}{2})2\pi\delta(\omega+\omega'),\\
\left\langle \tilde{x}_{in}(\omega)\tilde{y}_{in}(\omega')\right\rangle  & = & -\left\langle \tilde{y}_{in}(\omega)\tilde{x}_{in}(\omega')\right\rangle \\
 & = & \frac{i}{2}2\pi\delta(\omega+\omega'),
\end{array}\label{eq:noise sources correlation}
\end{equation}
and are given by

\begin{equation}
\begin{array}{ccc}
S_{\tilde{q}}(\omega) & = & [A_{1}(\omega)A_{1}(-\omega)+B_{1}(\omega)B_{1}(-\omega)](n_{a}+\frac{1}{2})\\
 &  & +[E_{1}(\omega)E_{1}(-\omega)+F_{1}(\omega)F_{1}(-\omega)](n_{m}+\frac{1}{2}),\\
S_{\tilde{p}}(\omega) & = & [A_{2}(\omega)A_{2}(-\omega)+B_{2}(\omega)B_{2}(-\omega)](n_{a}+\frac{1}{2})\\
 &  & +[E_{2}(\omega)E_{2}(-\omega)+F_{2}(\omega)F_{2}(-\omega)](n_{m}+\frac{1}{2}),
\end{array}\label{eq:power spectrum-1}
\end{equation}
where the first term proportional to $(n_{a}+\frac{1}{2})$ and the
second term proportional to $(n_{m}+\frac{1}{2})$ correspond to the
radiation pressure contribution and thermal noise contribution, respectively.
For $g_{0}=g_{1}=0$, Eq. (\ref{eq:power spectrum-1}) $S_{\tilde{q}}(\omega)=S_{\tilde{p}}(\omega)=\frac{\gamma_{m}}{\frac{\gamma_{m}^{2}}{4}+\omega^{2}}(n_{m}+\frac{1}{2})$
are simply Lorentzian lines with full width $\gamma_{m}$ at half
maximum. The variances in the position $\left\langle \delta\tilde{q}^{2}\right\rangle $
and momentum $\left\langle \delta\tilde{p}^{2}\right\rangle $ of
the mechanical mode are finally obtained by

\begin{equation}
\begin{array}{ccc}
\left\langle \delta Z{}^{2}\right\rangle  & = & \frac{1}{2\pi}\int_{-\infty}^{+\infty}S_{Z}(\omega)d\omega,\end{array}\label{eq:qq_pp_inter}
\end{equation}
giving rise to the numerical results in Fig. 2(c)-(d) (marker).


\begin{thebibliography}{10}
\bibitem{Lecocq_PRX2015}F. Lecocq, J. B. Clark, R. W. Simmonds, J.
Aumentado, and J. D. Teufel, Phys. Rev. X \textbf{5}, 041037 (2015).

\bibitem{Pirkkalainen_PRL2015}J. M. Pirkkalainen, E. Damsk\"agg, M.
Brandt, F. Massel, and M. A. Sillanp\"a\"a, Phys. Rev. Lett. \textbf{115},
243601 (2015).

\bibitem{Pontin_PRL2014}A. Pontin, M. Bonaldi, A. Borrielli, F. S.
Cataliotti, F. Marino, G. A. Prodi, E. Serra, and F. Marin, Phys.
Rev. Lett. \textbf{112}, 023601 (2014).

\bibitem{Szorkovszky_PRL2013}A. Szorkovszky, G. A. Brawley, A. C.
Doherty, and W. P. Bowen, Phys. Rev. Lett. \textbf{110}, 184301 (2013).

\bibitem{Schwab_NatPhy2009}J. B. Hertzberg, T. Rocheleau, T. Ndukum,
M. Savva, A. A. Clerk, and K. C. Schwab, Nat. Phys. \textbf{6}, 213
(2009). 

\bibitem{Ruskov_prb2005}R. Ruskov, K. Schwab, and A. N. Korotkov,
Phys. Rev. B \textbf{71}, 235407 (2005). 

\bibitem{Rabl_PRL2004}P. Rabl, A. Shnirman, and P. Zoller, Phys.
Rev. B \textbf{70}, 205304 (2004). 

\bibitem{Vanner_PNAS2011}M. R. Vanner, I. Pikovski, G. D. Cole, M.
S. Kim, C. Brukner, K. Hammerer, G. J. Milburn, and M. Aspelmeyer,
Proc. Natl. Acad. Sci. USA \textbf{108}, 16182 (2011).

\bibitem{Vanner_NatCom2013}{]} M. R. Vanner, J. Hofer, G. D. Cole,
and M. Aspelmeyer, Nat. Commun. \textbf{4}, 2295 (2013).

\bibitem{Aspelmeyer_PT2012}M. Aspelmeyer, P. Meystre, and K. Schwab,
Phys. Today \textbf{65}, 29 (2012).

\bibitem{Zurek_QCBoundary}W. H. Zurek, Phys. Today \textbf{44}, 36
(1991).

\bibitem{HighP_PRD1979}J. N. Hollenhorst, Phys. Rev. D \textbf{19},
1669 (1979).

\bibitem{Peano_PRL2015}V. Peano, H. G. L. Schwefel, Ch. Marquardt,
and F. Marquardt, Phys. Rev. Lett. \textbf{115}, 243603 (2015).

\bibitem{Rugar_PRL1991_3db_limit}D. Rugar and P. Gr\"utter, Phys. Rev.
Lett. \textbf{67}, 699 (1991). 

\bibitem{Aspelmeyer_RMP2014}M. Aspelmeyer, T. J. Kippenberg, and
F. Marquardt, Rev. Mod. Phys. \textbf{86}, 1391 (2014).

\bibitem{Clerk_njp2008}A. A. Clerk, F. Marquardt, and K. Jacobs,
New J. Phys. \textbf{10}, 095010 (2008).

\bibitem{Zhang_pra2009}J. Zhang, Y. X. Liu, and F. Nori, Phys. Rev.
A \textbf{79}, 052102 (2009). 

\bibitem{Gu_pra2013}W. J. Gu, G. X. Li, and Y. P. Yang, Phys. Rev. A
\textbf{88}, 013835 (2013). 

\bibitem{Asjad_pra2014}M. Asjad, G. S. Agarwal, M. S. Kim, P. Tombesi,
G. DiGiuseppe, and D. Vitali, Phys. Rev. A \textbf{89}, 023849 (2014).

\bibitem{Lv_pra2015}X. Y. L\"u, J. Q. Liao, L. Tian, and F. Nori, Phys.
Rev. A \textbf{91}, 013834 (2015). 

\bibitem{Wang_PRL2013}Y. D. Wang and A. A. Clerk, Phys. Rev. Lett.\textbf{
110}, 253601 (2013).

\bibitem{Jahne_pra2009}K. J\"ahne, C. Genes, K. Hammerer, M. Wallquist,
E. S. Polzik, and P. Zoller, Phys. Rev. A \textbf{79}, 063819 (2009).

\bibitem{Huang_pra2010}S. M. Huang and G. S. Agarwal, Phys. Rev.
A \textbf{82}, 033811 (2010).

\bibitem{Kron_pra2013}A. Kronwald, F. Marquardt, and A. A. Clerk,
Phys. Rev. A \textbf{88}, 063833 (2013).

\bibitem{Woll_Sci2015}E. E. Wollman, C. U. Lei, A. J. Weinstein,
J. Suh, A. Kronwald, F. Marquardt, A. A. Clerk, and K. C. Schwab,
Science \textbf{349}, 952 (2015).

\bibitem{Lei_PRL2016_3db}C. U. Lei, A. J. Weinstein, J. Suh, E. E.
Wollman, A. Kronwald, F. Marquardt, A. A. Clerk, and K. C. Schwab,
Phys. Rev. Lett. \textbf{117}, 100801 (2016). 

\bibitem{Mari_PRL2009}A. Mari and J. Eisert, Phys. Rev. Lett. \textbf{103},
213603 (2009).

\bibitem{Pinaotey_pra2014} S. Pina-Otey, F. Jimenez, P. Degenfeld-Schonburg,
and C. Navarrete-Benlloch, Phys. Rev. A \textbf{93}, 033835 (2016).

\bibitem{Agarwal_pra2016}G. S. Agarwal and S. M. Huang, Phys. Rev.
A \textbf{93}, 043844 (2016).

\bibitem{Xuereb_pra2012}A. Xuereb, M. Barbieri, and M. Paternostro,
Phys. Rev. A \textbf{86}, 013809 (2012).

\bibitem{Farace_pra2012}A. Farace and V. Giovannetti, Phys. Rev.
A \textbf{86}, 013820 (2012).

\bibitem{LvXY_PRL2015}X. Y. L\"u, Y. Wu, J. R. Johansson, H. Jing,
J. Zhang, and F. Nori, Phys. Rev. Lett. \textbf{114}, 093602 (2015).

\bibitem{Walls_book2007}D. F. Walls and G. J. Milburn, Quantum Optics
(Springer, Berlin, 2008).

\bibitem{NOTE} The pump field with the frequency being twice of the
cavity resonance can in principle co-propagate with the amplitude-modulated
driving field with high transmission through the fixed end mirror
of a designed coating.

\bibitem{Gardiner_book2004}C. W. Gardiner and P. Zoller, Quantum
Noise, 3rd ed. (Springer, New York, 2004).

\bibitem{Giovannetti_pra2001}V. Giovannetti and D. Vitali, Phys.
Rev. A \textbf{63}, 023812 (2001).

\bibitem{Clerk_RMP2010}A. A. Clerk, M. H. Devoret, S. M. Girvin,
F. Marquardt, and R. J. Schoelkopf, Rev. Mod. Phys. \textbf{82}, 1155
(2010).

\bibitem{Benguria_PRL1981}R. Benguria and M. Kac, Phys. Rev. Lett.
\textbf{46}, 1 (1981).

\bibitem{Vitali_pra2007}D. Vitali, P. Tombesi, M. J. Woolley, A.
C. Doherty, and G. J. Milburn, Phys. Rev. A \textbf{76}, 042336 (2007).

\bibitem{Groblacher_NatPhys2009}S. Gröblacher, J. B. Hertzberg, M.
R. Vanner, S. Gigan, K. C. Schwab, and M. Aspelmeyer, Nat. Phys. \textbf{5},
485 (2009).

\bibitem{Ludwing_njp2008}M. Ludwig, B. Kubala, and F. Marquardt,
New J. Phys. \textbf{10}, 095013 (2008).

\bibitem{Mari_njp2012}A. Mari and J. Eisert, New J. Phys. \textbf{14},
075014 (2012).

\bibitem{Gigan_Nature2006}S. Gigan, H. Böhm, M. Paternostro, F. Blaser,
G. Langer, J. Hertzberg, K. Schwab, D. Bäuerle, M. Aspelmeyer, and
A. Zeilinger, Nature (London) \textbf{444}, 67 2006.

\bibitem{Kleckner_Nature2006}D. Kleckner and D. Bouwmeester, Nature
(London) \textbf{444}, 75 (2006).

\bibitem{Marquardt_PRL2007}F. Marquardt, J. P. Chen, A. A. Clerk,
and S. M. Girvin, Phys. Rev. Lett. \textbf{99}, 093902 (2007).

\bibitem{Schonburg_pra2016}P. Degenfeld-Schonburg, M. Abdi, M. J.
Hartmann, and C. Navarrete-Benlloch, Phys. Rev. A \textbf{93}, 023819
(2016).

\bibitem{pra1987_Dejesus}E. X. DeJesus and C. Kaufman, Phys. Rev.
A \textbf{35}, 5288 (1987).

\bibitem{RMP2012_Weed}C. Weedbrook, S. Pirandola, R. García-Patrón,
N. J. Cerf, T. C. Ralph, J. H. Shapiro, and S. Lloyd, Rev. Mod. Phys.
\textbf{84}, 621 (2012).

\bibitem{www_Teschl}G. Teschl, Ordinary Differential Equations and
Dynamical Systems, http://www.mat.univie.ac.at/gerald.
\end{thebibliography}
\end{document}